\documentclass[aps,pre,showpcs,amsmath,amssymb,superscriptaddress,reprint]{revtex4-1}

\usepackage{graphicx}
\usepackage{caption}
\usepackage{subcaption}
\usepackage{dcolumn}
\usepackage{bm}
\usepackage{hyperref}

\begin{document}
\author{Sanjin Marion}
\author{Michal Macha}
\author{Sebastian J. Davis}
\author{Andrey Chernev}
\author{Aleksandra Radenovic}
\email{aleksandra.radenovic@epfl.ch}
\affiliation{Laboratory of Nanoscale Biology, Institute of Bioengineering, School of Engineering, EPFL, 1015 Lausanne, Switzerland}

\title[]{Wetting of nanopores probed with pressure}

\title[]{Wetting of nanopores probed with pressure}

\keywords{nanopore, nanobubble, pressure, 2D materials, hydrophobic, molybdenum disulfide }




\begin{abstract}
Nanopores are both a tool to study single-molecule biophysics and nanoscale ion transport,  but also a promising material for desalination or osmotic power generation. Understanding the physics underlying ion transport through nano-sized pores allows better design of porous membrane materials. Material surfaces can present hydrophobicity, a property which can make them prone to formation of surface nanobubbles. Nanobubbles can influence the electrical transport properties of such devices. We demonstrate an approach which uses hydraulic pressure to probe the electrical transport properties of solid state nanopores. We show how pressure can be used to wet pores, and how it allows control over bubbles in the nanometer scale range normally unachievable using only an electrical driving force. Molybdenum disulfide is then used as a typical example of a 2D material on which we demonstrate wetting and bubble induced nonlinear and linear conductance in the regimes typically used with these experiments. We show that by using pressure one can identify and evade wetting artifacts.
\end{abstract}
\maketitle

\section{Introduction}

A solid state nanopore consists of a solid membrane dividing two solute containers with an opening (pore) with a diameter in the range ~1-100 nm\cite{Dekker2007}. By measuring the changes in the conductance of the sample as a particle passes through the pore opening the particle's properties, such as size, shape, and charge, can be determined with great accuracy in as low as atto-molar analyte concentrations\cite{Wanunu2010}. When the diameter of nanopores is reduced to tens of nanometers and below, they can become selective to ions\cite{Vlassiouk2008} and can provide information on the effective diffusion constants of different ion species\cite{Rollings2016}.  Nanopores, and related nanochannels, also enable the study of physical phenomena on the nanoscale\cite{Schoch2008,Bocquet2010,Mouterde2019} and can be used as building blocks for various nanoscale devices which can act as ion pumps\cite{Siwy2002}, electrical diodes\cite{Siwy2005}, desalination membranes\cite{Werber2016}, or osmotic power generation devices\cite{Macha2019}. Recently, 2D materials like graphene\cite{Garaj2010,Schneider2010b,Merchant2010}, hexagonal boron nitride (h-BN)\cite{Zhou2013}, and molybdenum disulfide (MoS$_2$)\cite{Liu2014} have become a popular nanopore membrane because their thickness is comparable to the size of single nucleotides or ions. Due to the ease of nanopore fabrication they are readily used to probe  nonlinear ion transport phenomena\cite{Jain2015, Feng2016, Thiruraman2018}.

Improper wetting and cleaning of nanopore membranes has been connected to an increase in the noise level during ionic current measurements\cite{Smeets2008,Uram2008,Beamish2012}. Through localized laser heating it was suggested that bubbles in the nm-size range can remain in nanopores and reduce the conductance of the pores while increasing the electrical noise level\cite{Smeets2006a}. Solid-state nanopores have also been used to nucleate nanobubbles through liquid super-heating by large electric fields\cite{Nagashima2014,Levine2016} or with plasmonic nanopores\cite{Li2015a}.  Although bulk gas nanobubbles were identified nearly 20 years ago\cite{Lou2000,Ishida2000}, their long term stability on surfaces was a matter of controversy until recently\cite{Alheshibri2016}. Nanobbubles have been demonstrated to be stable for days or even months on surfaces with receding contact angles larger than $80^\circ$ or advancing contact angles of $70^\circ$ or higher (e.g. Mica, HOPG, coated Si)\cite{Lohse2015}. A hydrophobic surface is defined by a receding contact angle above $90^\circ$\cite{Law2014}, so even mildly hydrophobic surfaces can support stable surface nanobubbles. 

Nanobubbles made from ambient air have been detected on graphene surfaces using transmission electron microscopy\cite{Shin2015a} and atomic force micrscopy\cite{Cantley2019}.  We expect 2D materials to have two properties which would contribute to the stability of surface nanobubbles: the presence of contaminants during material transfer\cite{Graf2019}, and an inherently low wettability. 2D materials can have a large variation in the degree of hydrophobicity depending on material type, defect density, and sample quality, with contact angles above $70^\circ$ reported for MoS$_2$\cite{Gaur2014,Chow2015,Annamalai2016}, h-BN\cite{Li2017b}, and graphene\cite{Taherian2013,Annamalai2016}. In addition, sample defect density and ambient air exposure has been shown to increase the hydrophobicity of these materials, with fresh samples being more hydrophilic\cite{Kozbial2014,Chow2015}. Some of these materials, like h-BN\cite{Zhou2013}, can be made more hydrophilic by chemical or ozone treatments.

Identifying the presence of nanobubbles and wetting issues on nanopores can be a daunting task. If only ion transport measurements are used, it provides limited information on the nature of the conductance pathway through a pore and needs to be coupled with modeling and prior imaging of pores. Our approach to provide additional information is to use hydraulic pressure as a complementary probe to ionic transport. Hydraulic pressure is known to modify ionic current rectification of nanopores\cite{Lan2011}, and can be used to probe the role of fluid flow in nanopore electrical transport\cite{Jubin2018}, or fluid flow on the angstrom scale\cite{Mouterde2019}. The use of pressure with thin nanopores has mostly been oriented towards influencing DNA or protein translocations \cite{Buyukdagli2015,Li2017a,Zhang2018} or for characterising the role of pore surface properties for translocations\cite{Firnkes2010,Waduge2017}. For ultrathin membrane materials as the ones used in this work, the flow through nanopores is dominated by hydrodynamic access resistance at pore entrances\cite{Mao2014,Gravelle2014}, similar to conductance effects. The nature of the flow lines is that they converge at pore entrances  which can drag objects through the pore\cite{Lan2011a,Gadaleta2015}. Additionally, hydrostatic pressure has been demonstrated as a way to wet hydrophobic porous materials\cite{Smirnov2010}, an alternative to electrical field induced wetting\cite{Smirnov2011,Powell2011}, but has not yet been demonstrated on single pores. 

The aim of this work is to demonstrate that nanoscale bubbles originating from dissolved ambient air in the solutions can significantly modify the measured ionic transport properties of nanopores. First, we will demonstrate the combination of pressure and voltage probes and use it to control the wetting state of a single hydrophobic nanopore. This allows us to establish a baseline of how improper wetting and its effect on ionic conductance of a pore manifests. Secondly, we show how pressure provides a means of identifying and removing nanobubbles or other wetting artifacts in hydrophilic nanopores.  Lastly, we will study bubble and wetting induced ionic transport artifacts on MoS$_2$ and demonstrate how alcohol prewetting can grow nanobubbles. Pressure is demonstrated as a viable alternative for wetting and probing the state of nanopores.

\section{Nanofluidics set-up description}

\begin{figure*}
\centering
\includegraphics{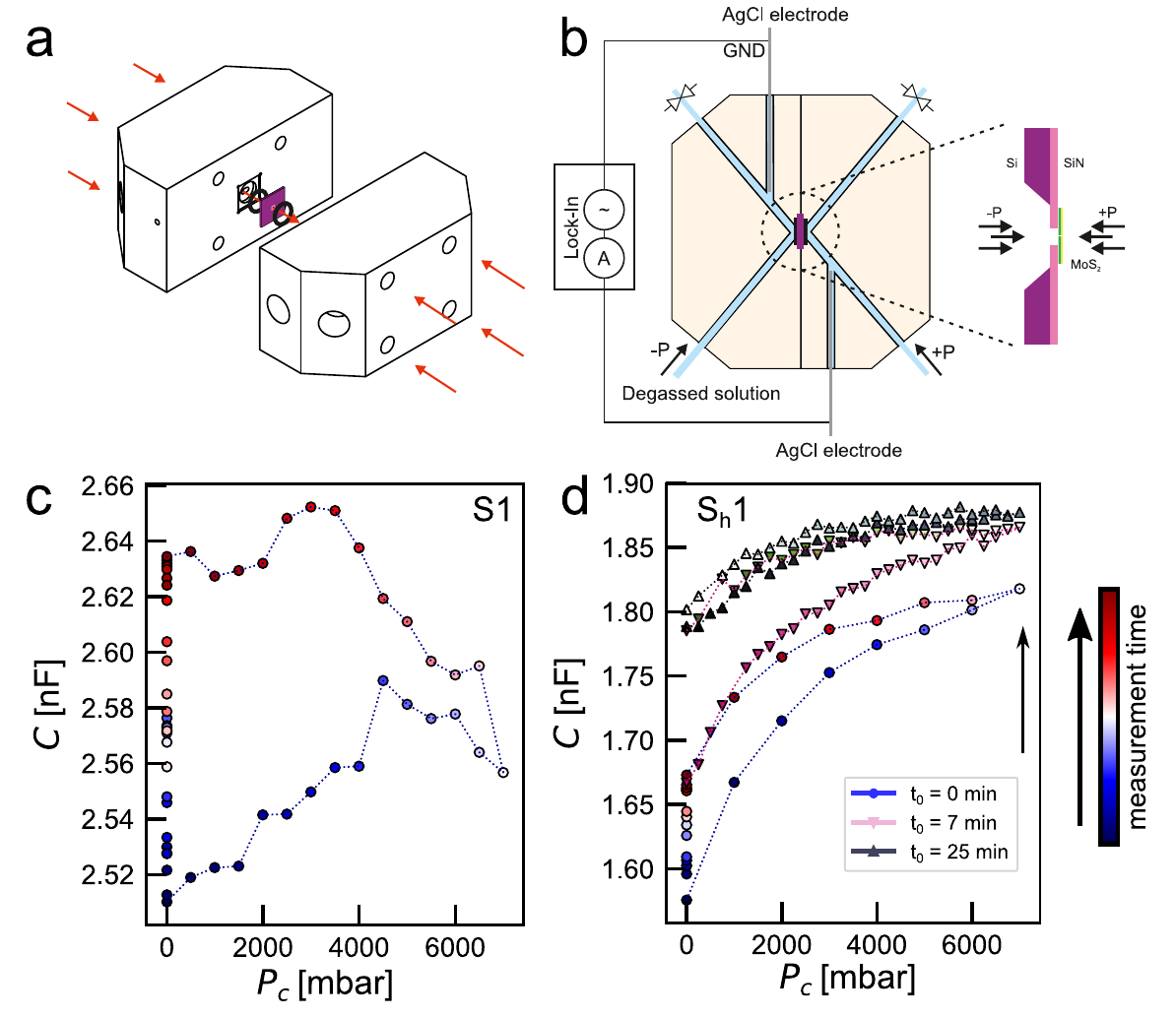}
\caption{\small \textbf{Applying pressure to solid-state nanopores} a) Diagram of the sample chamber b) Simplified diagram of the fluidic and electrical connections. c) Capacitance of a partially wetted supported membrane of $\approx$20x20 $\mu$m$^2$ and 20nm thickness as a function of applied compression pressure during the wetting process. The pressure $P$ is swept from 0 to 7 bar, with intermediate steps at $P=0$ to check for changes in the baseline value of the pressure. The solution was 1 M KCl Tris buffered to pH8. Colours of the points go from blue (first measurement) to red (last measurement) in a continuous fashion, as marked by the arrow. The measurement took about 5 minutes. Ion transport measurements through the pore in this sample are shown on Fig.\ \ref{fig:bubble} and in the Supporting information. d) The same as in panel c) except the chip has a smaller supported membrane size of $\approx$12x12 $\mu$m$^2$ and the chip was coated with a hexamethyldisilazane (HMDS) before use to make the surface more hydrophobic and was not cleaned in any way afterwards. A series of three consecutive measurements are provided with start times at $t_0$. The first measurement took about 5 min, the last two lasted for about 10 min.}
\label{fig:setup}
\end{figure*}

Applying hydrostatic pressure on membranes with nanopores requires water-tight sealing of the sample. This has the added benefit of minimizing water evaporation and gas permeation. To that end, we designed a closed microfluidics chamber made from PEEK and PTFE components which sustain pressures in excess of 10 bar (Fig.\ \ref{fig:setup}a). The sample is placed between the two halves of the chamber and sealed using two nitrile O-rings. On each side of the chip there are three fluid pathways: one for an electrode, one for a fluid inlet, and one for a fluid outlet. On both outlets exchange reservoirs are placed onto which compressed gas with pressures up to 7 bar is applied using a microfluidics pressure controller which is also used to measure the applied pressure $P$. The design has minimal crevices and corners to minimize bubble formation during filling of the chambers\cite{Pereiro2019}. Filling of the chamber is done with solutions which have been degassed from dissolved ambient air  in the connecting fluid pathway using a degassing tube. We find that any mixing of the solution or contact with ambient air results in significant gas absorption, regassing the liquids.

By applying a voltage bias between the two Ag/AgCl electrodes on different sides of the membrane we are able to measure current passing through the nanopore (Fig.\ \ref{fig:setup}b). In addition, we apply a sinusoidal set voltage (AC) and measure the corresponding current using a phase sensitive amplifier (lock-in). This allows the use of the delay in the AC signal to calculate the capacitance of the membrane. We assume the simplest case of a parallel connection of a capacitance $C$ and resistor $R_{ac}$, where the resistance $R_{ac}$ corresponds to the resistance of the pore $R$ as obtained at DC values when the frequency of the AC signal is kept at around 1 Hz (See Supplemental information Sec. S2). In reality the frequency response of a membrane has other contributions but they are only relevant at higher frequencies\cite{Dimitrov2010,Traversi2013}, with a flat plateau at sufficiently low frequencies where measurements are conducted. For 20nm thick suspended silicon nitride membranes in the size ranges of 10x10 to 50x50 $\mu$m$^2$ which we use here we obtain the expected values for the capacitance on the order of $\approx$1 nF (in 1M KCl Tris buffered to pH8)\cite{Dimitrov2010} and proportional to the supported membrane area.

The microfluidics chamber allows the application of hydrostatic pressure in a gradient condition $\Delta P$, with a positive pressure gradient defined as pressure applied from the front side of the membrane, and negative from the back side (Fig.\ \ref{fig:setup}b). It also allows applying compression pressure $P_c$ on both sides of the membrane simultaneously. Applying pressure on one or both sides of the chip is used as an additional probe of the system. We alternate between no applied pressure $P=0$ and a cycle of pressures $P_i$ going from $P_0=0$ to $P_i= P_{max}$, and back down to $P_i=0$. In the case that a gradient of pressure is applied, negative pressures are also applied following the same procedure. This allows the detection of any hysteresis in a measured response to the pressure and if the base value at $P=0$ is changing during the measurement. During the pressure cycle either the current response to a bias voltage or the resistance and capacitance using an AC response are measured. The representative value is measured after a sufficient settling time. Examples of applied pressures and measured responses are provided in the Supplemental information.

Compression of solutions, especially degassed ones, has been shown to be comparable in effectiveness to alcohol wetting\cite{Pereiro2019}. To demonstrate this and to prove proper wetting of membranes, the amount of surface area of the membrane in contact with the liquid is monitored via the membrane capacitance $C$. An example measurement of capacitance for several sweeps of compression pressure is shown on Fig.\ \ref{fig:setup}c,d for a hydrophilic (Sample $S$1) and a hydrophobic surface (Sample $S_h$1). As pressure is applied to a freshly degassed solution the baseline value of C, at intermediary $P=0$ steps in the sweep protocol, increases throughout the protocol. In the case of solid state nanopores the largest contribution to the capacitance at low frequencies is known to come from the suspended membrane itself\cite{Dimitrov2010}. If part of the suspended membrane with a surface area $\Delta A$ is not wetted, i.e. there are bubbles or air patches present, then the total capacitance of the membrane will be reduced by an amount $\Delta C \sim \Delta A$ (See Supplemental Material Sec.\ {S3 for details). In that sense, the increase of capacitance as pressure is applied is interpreted as the membrane being wetted and gas from the bubbles being absorbed into the liquid which was under-saturated with gas. Hydrophilic membranes fill fast, within tens of seconds of applied compression pressure with a degassed solution, while hydrophobic pores sometimes seem to have remnants of bubbles left on the surface (Fig.\ \ref{fig:setup}d.  In contrast if a non-degassed solution is used, we report that the capacitance often returns close to its original value, implying that if the solution is saturated excess dissolved gas can return to the membrane in the form of gas bubbles, or that the bubbles are impervious to absorption. In all the following measurements the membranes were flushed with degassed solutions and compression pressures of up to 7 bar were applied for $\approx$5 minutes to ensure proper wetting.

\section{Wetting nanopores coated with hydrophobic polymers}

\begin{figure*}
\centering
\includegraphics{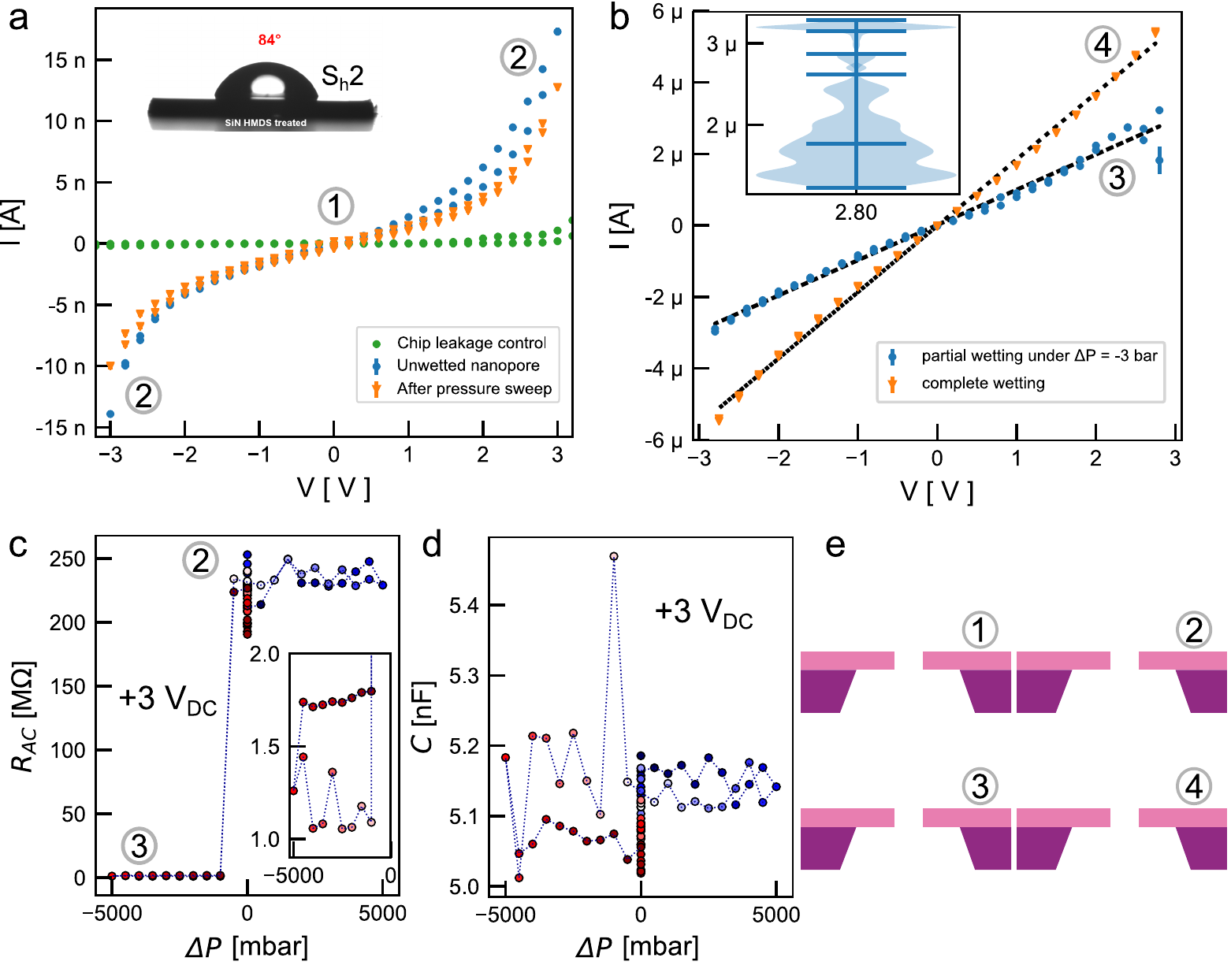}
\caption{\small \textbf{Pressure and voltage wetting of a surface treated nanopore (Sample S$_h$2).} 
a) IV curve in the un-wetted state for a $~130$ nm diameter nanopore in 20 nm thick silicon nitride coated with HMDS in 1 M KCl Tris buffered to pH 8. Green circles show the leakage current from a control membrane without a pore. Blue circles show the measured current after initial wetting. Orange triangles correspond to a measurement done after the experiments in panels b) and c). At low voltages ($<0.1$ V) the unwetted pore has a finite resistance of $\approx$1 G$\Omega$. Inset shows the contact angle measurement with a macroscopic water droplet on a silicon nitride chip coated with HDMS.
b) Blue circles are an IV curve for the same sample as in panel a) but after several sweeps of $\pm 10$ V and recorded under $\Delta P=-3$ bar of constant pressure. The dashed line is a fit indicating a resistance $R=0.99$ M$\Omega$ which is larger than the expected resistance $R\approx 0.5$ M$\Omega$ calculated based on the size. Inset shows the increase in the fluctuations at high voltages where the bars mark the extrema of the individual current traces and the shape of the curves denotes the probability density of having the current at a certain value during the measurement. The orange triangles show an IV curve after complete electro-wetting where the pore shows linear behaviour, no fluctuations, and a correct value of the resistance. Note that the orange curve was obtained after all the different wetting states were characterised.
c) Pressure sweep performed after panel b) under a DC bias of +3 V with the resistance and capacitance measured using a superimposed 100 mV AC bias. Inset shows a zoomed view of the negative pressure values. 
d) Capacitance measurement obtained simultaneously as the data in panel c). 
e) Schematic diagram of an unwetted pore based on the electrowetting model\cite{Smirnov2011}: (1), menisci overlap under pressure and/or voltage (2), partially wetted (3) and fully wetted nanopore (4). Numbers in panels a), b), and c) correspond to the interpreted states of the nanopore.}
\label{fig:electrowetting}
\end{figure*}

We will first demonstrate how the combination of electrical and pressure probes can be used to wet hydrophobic nanopores. To that end a 130 nm diameter pore in 20 nm thick silicon nitride was coated with a hydrophobic silane hexamethyldisilazane (HMDS) so that we assume that the interior of the pore is also covered. The final static contact angle of  $\approx 84^\circ$ was measured by a macroscopic water drop on the chip. The nanopore was wetted using a degassed solution of 1 M KCl Tris buffered to pH8 in the chamber. Fig.\ \ref{fig:electrowetting}a shows the measured current versus applied voltage bias through the membrane after wetting for sample S$_h$2. The measured dependence of current $I$ versus applied DC voltage $V$ (IV curve) has a characteristic nonlinear shape which is not related to the leakage current through the membrane. Although high voltages can cause pore enlargement or formation, for the material used here the dielectric breakdown voltage is expected to be $\approx 10$ V but will highly depend on the silicone nitride properties such as defect density.\cite{Kwok2014} To ensure that the measured IV curves are not the result of a leakage current through the membrane, the same measurements on membranes from the same production batch but without any nanopores are performed (Fig.\ \ref{fig:electrowetting}a), indicating leakage current is negligible. Wetting curves such as the one in Fig.\ \ref{fig:electrowetting}a were obtained in a study on voltage-gating in hydrophobic nanopores\cite{Smirnov2011}, where they were attributed to the electrical field forcing the gas-liquid interface menisci on opposite sides of the membrane to touch through the nanopore. This produces a nonlinear voltage dependent conductance as higher voltages increase the overlap area of the menisci. Applying higher potential differences between the two sides of the nanopore can wet the pore in a process termed electro-wetting even at no applied pressure difference\cite{Smirnov2011,Beamish2012,Beamish2012}. This state was found to be temporary, reverting to an unwetted state after some time, unless even higher voltages were applied which would drive all the remnants of gas out of the pore area and remove any gas nucleation sites.

In order to transition the sample from its unwetted state in Fig.\ \ref{fig:electrowetting}a to a wetted one a combination of pressure gradients and DC voltage bias are applied. After sweeping the applied DC voltage an application of a $\Delta P = -3$ bar was sufficient to wet the pore (Fig.\ \ref{fig:electrowetting}b). Although pressure gradient induced wetting of nanopores was described and measured in porous media\cite{Smirnov2010}, to our knowledge this is the first instance of pressure induced wetting of a single nanopore. The current baseline of this wetted state was stable except around +3 V of bias which demonstrated fluctuations in the current level. These fluctuations have been detected previously near critical voltages for electrowetting\cite{Smirnov2011}, and we attribute them here to partial wetting/de-wetting as already described in both solid state pores\cite{Smirnov2011,Powell2011} and biological channels\cite{Aryal2015}. If a sweep of pressure ($0$ bar $\rightarrow$ 7 bar $\rightarrow -7$ bar $\rightarrow 0$ bar) was performed, the state of the pore changed reversibly from the unwetted to the wetted and back to the unwetted state (Fig.\ \ref{fig:electrowetting}c). During this pressure sweep there was no significant change of capacitance (Fig.\ \ref{fig:electrowetting}d) which we interpret as no large bubbles moving on/off the membrane and changing its capacitance. An IV curve measured after this switching (Fig.\ \ref{fig:electrowetting}a) closely matches the IV curve of the original unwetted state. Note that the resistance in the partially wetted state is still larger than the expected resistance for these pores by a factor of two. We interpret this as the pore interior not being completely wetted, in accordance to the scenario proposed that permanent and complete electrowetting is achieved with applying higher voltages\cite{Smirnov2011}. After further application of high voltage and compression pressure the pore was wetted with a resistance of $\approx 0.5$ M$\Omega$ matching the expected resistance for a nanopore of this size (Fig.\ \ref{fig:electrowetting}b). In an ideal case one would expect that both positive and negative pressure should wet the pore, but an asymmetry in the nanopore unwetted volume is attributed to a preference for one direction of the pressure gradient. In conclusion, pressure and electrical potential bias can be used to wet even hydrophobic pores. Attempts to wet smaller coated pores ($\approx$75 nm) were unsuccessful, consistent with previous studies indicating that even higher critical voltages and pressures would be required\cite{Smirnov2011}. Sample S$_h$1 was wetted with compression pressure after several hours of high voltage IV curves, but its coating was less successful with a measured angle of $\approx 78^\circ$ (Supplemental information Sec.\ S5).

\section{Wetting artifacts in hydrophilic pores }

\begin{figure*}
\centering
\includegraphics{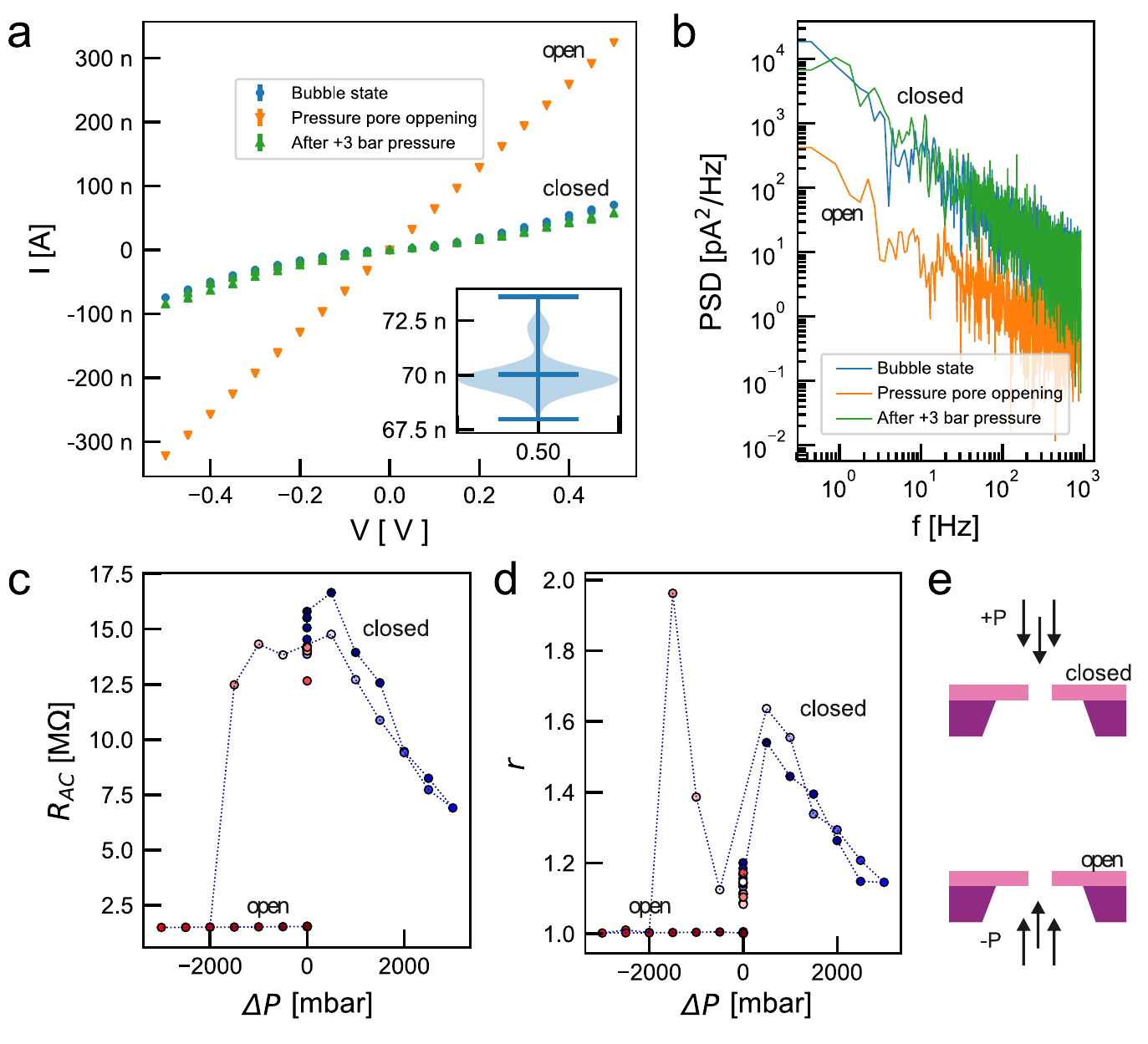}
\caption{\small \textbf{Pressure gatting of nanobubbles in hydrophylic nanopores (Sample S1).} a) IV curve of a bubble inside a $d \approx 75$ nm diameter nanopore before applying pressure (blue), after opening with pressure (orange), and after closing with pressure using +3 bar (green). The resistance of the open pore is found to be $R=1.61$ M$\Omega$ which is within 10\% of the expected resistance of the nanopore in 1 M KCl Tris buffered to pH8. Inset shows the extrema and probability density for current fluctuations at $V=+0.5$ V for the bubble state.  b) Power spectral densities of the current traces at +0.2 V from panel a). c) Resistance obtained during a pressure sweep performed on the pore from panel a) which switched the pore from an obstructed to a open state. A time trace of this measurement along with the capacitance $C$ is provided in the Supplemental information. d) Rectification $r(V)= |I(+0.1 V) | / |I(-0.1V)|$ as measured during the pressure sweep in panel c). e) Proposed toy model of the hypothesized nanobubble wetting behavior with different applied pressure gradients. See main text for an in depth discussion.}
\label{fig:bubble}
\end{figure*}

Even with hydrophilic nanopores one can obtain temporary obstructions or other unexpected phenomena which are hard to interpret. In the process of testing our set-up we have found that improper degassing of the solutions in use increases the likelihood of wetting issues. We used fresh oxygen plasma treated silicon nitride pores with a diameter of $d\approx 75$ nm which are standard support membranes for working with 2D materials. It was found that wetting artifacts would be induced at the nanopore under an applied voltage or pressure gradient at the start or during measurements. An example of a measured IV curve we attribute to a nanobubble being pinned at the pore entrance is shown on Fig.\ \ref{fig:bubble}a. The IV curve presents a nonlinear shape similar to the wetting curve in Fig.\ \ref{fig:electrowetting} albeit with ten times higher currents at the same voltage. It was found that pressure gradients or even higher voltages could be used to change the state of the pore. By sweeping the gradient pressure in the range of $\Delta P=\pm 3$ bar the state of the pore was changed from a resistive and nonlinear one ($\approx 15$ M$\Omega$) to a linear and more conductive one ($R\approx 1.6$ M$\Omega$). An IV curve corresponding to this linear and conductive state is shown in Fig.\ \ref{fig:bubble}a which confirms the stability of such a state for minutes or longer.  Then by applying $\Delta P= +3$ bar for several seconds the resistance is again returned into a similar obstructed state. We argue that the pressure induced fluid flow is moving an object in and out of the pore, partially obstructing it. Any flow through a nanopore will have the flow lines bent and converging towards the pore entrance producing drag forces. The resulting drag force will have components in both the plane of the membrane and in the direction through the pore, which can then cause both translocations and moving of objects closer to the pore center. The difference between hydraulic pressure induced flow and electrical field induced drag (which would also include electroosmotic flow\cite{Melnikov2017}) is in the streamline shape and the magnitude of the drag force. The magnitude of pressure induced drag is proportional to the dimensions of the object while the electric field induced force is additionally dependent on the surface charge.

The reason why the behavior on Fig.\ \ref{fig:bubble} is not attributed to a wetting-dewetting transition, but to a nanobubble obstruction, lies in the noise level, the dependence of the resistance on pressure, and the presence of ionic current rectification. The current power spectral density (Fig.\ \ref{fig:bubble}b) is consistent with previous works predicting that nanobubbles in solid state pores increase the noise level\cite{Smeets2006a}. The noise power spectral density at frequencies below several kHz is dominated by flicker noise, which is known to scale with frequency as $S\approx A  f^{-\beta}$ with $\beta\approx 0.5 -1.5$ and $A$ the flicker noise amplitude\cite{Tabard-Cossa2007,Uram2008,Gravelle2019}.  We find the flicker noise level in all the samples presenting bubble issues to be slightly higher than ideal for such samples\cite{Tabard-Cossa2007,Smeets2008,Uram2008}.  It has also been predicted to increase in the case of wetting issues\cite{Beamish2012}. We notice several pressure induced behaviors in our measurements. First, the noise would sometimes increase or decrease after applying a pressure gradient, probably depending on if the pore was wetted or dewetted with the pressure. Secondly, we note that in some cases a decrease of resistance after applying pressure was followed with an increase in flicker noise amplitude. We speculate that the nature of the flicker noise will most likely depend on the surface charge of the air bubble, its shape, and position within the pore. For example if the bubble is changing the resistance dominantly from obstructing the pore channel or the access area to the pore, the contributions to the flicker noise will be different\cite{Fragasso2019}, with predictions that the surface contribution is more pronounced than the bulk contribution in nanopores smaller than 20 nm. We conclude that the noise level can strongly vary based on the position and shape of the obstructing bubble, with the lowest possible noise level only achieved once the whole pore region is completely wetted\cite{Tabard-Cossa2007,Beamish2012}. The mechanism of bubble pining here is unclear, but from the ease of moving the bubbles between an obstructing and non-obstructing state we postulate that it could be due to small defects or contaminants that survived the sample cleaning procedure. It could also be the fact that the pore is too small to let a bulk nanobubble pass through it.

Fig.\ \ref{fig:bubble}c shows that in the closed state there is a decrease in resistance of the pore as a pressure gradient is applied from 0 to 3 bar. If the bubble was pinned on the front side of the membrane then once the bubble is inside the pore additional flow from increased pressure gradient would deform and elongate the bubble thus opening up a wider pathway for fluid flow and decreasing the resistance. Something that would not occur in the case of a solid obstruction. We notice that the obstructing object is easily flushed through by a longer and consistent use of higher voltages or pressures in either direction, making it hard to obtain these measurements. 
Another indicator connected to the presence of nanobubbles in solid state pores is the level of ionic current rectification\cite{Li2015a}. Ionic current rectification defined as $r(V)= |I(+V) | / |I(-V)|$ is known to increase when there is geometrical asymmetry or surface charge asymmetry in nanopores. Since our nanopores are symmetric, an increase in $r$ would be an additional indicator of a nanobubble or other obstructing object being present. The values of $r$ calculated from the AC response during the pressure sweep are provided in (Fig.\ \ref{fig:bubble}d). The ionic current rectification increases when the pore is in the obstructed state, and is practically nonexistent in the open state (See supplemental information for details). We performed finite element modeling using coupled Poisson-Nernst-Planck-Stokes equations of four different example static scenarios: a) an open pore, b) a pore with an air bubble next to the pore entrance, c) a pore with an obstructing object clogging the pore, and d) a pore constricted by a symmetric air bubble. In all cases, the change of resistance versus pressure is negligible. Pressure induced flow is found to reduce ionic current rectification in the case of large openings, but not in the case of small pores. This effect has been studied previously and was attributed to flow negating the local charge distribution in the nanopore which is responsible for ionic current rectification\cite{Lan2011, Jubin2018}.

In conclusion, we find that none of these four static cases can explain such a large change in the resistance R or rectification r versus pressure P (See Supplemental information Sec.\ S4 for the details), strengthening our hypothesis that flow is inducing a movement or deformation of the obstructing object. In addition, we were not able to obtain the activated non-linear IV curve in these models as seen in the case of the closed state in Fig.\ \ref{fig:bubble}. Previous works have either attributed this non-linearity to temporary electrowetting\cite{Smirnov2011} or to hydration layer shredding\cite{Cantley2019,Jain2015}. Hydration layer shredding would require that the constriction is comparable to the size of the ions, which is inconsistent with the large variaton of both resistance and rectification with pressure, especially as it is known that pressure does not reduce ionic current rectification for small nanopores due to too low flow rates\cite{Lan2011}. Here we note that surface conduction and not Debye layer overlap is responsible for ionic current rectification\cite{Poggioli2019}, so that due to the nanobubble surface charge the unobstructed area of the pore opening can have larger dimensions than the Debye screening length which is $\approx 0.3$ nm in our conditions. We conclude that the presented experimental data supports a dynamic change of the obstructing object at the nanopore which is consistent with bad wetting most likely due to the presence of a bubble in the nanoscale size range. 

\section{Wetting artifacts with 2D materials}

\begin{figure*}
\centering
\includegraphics{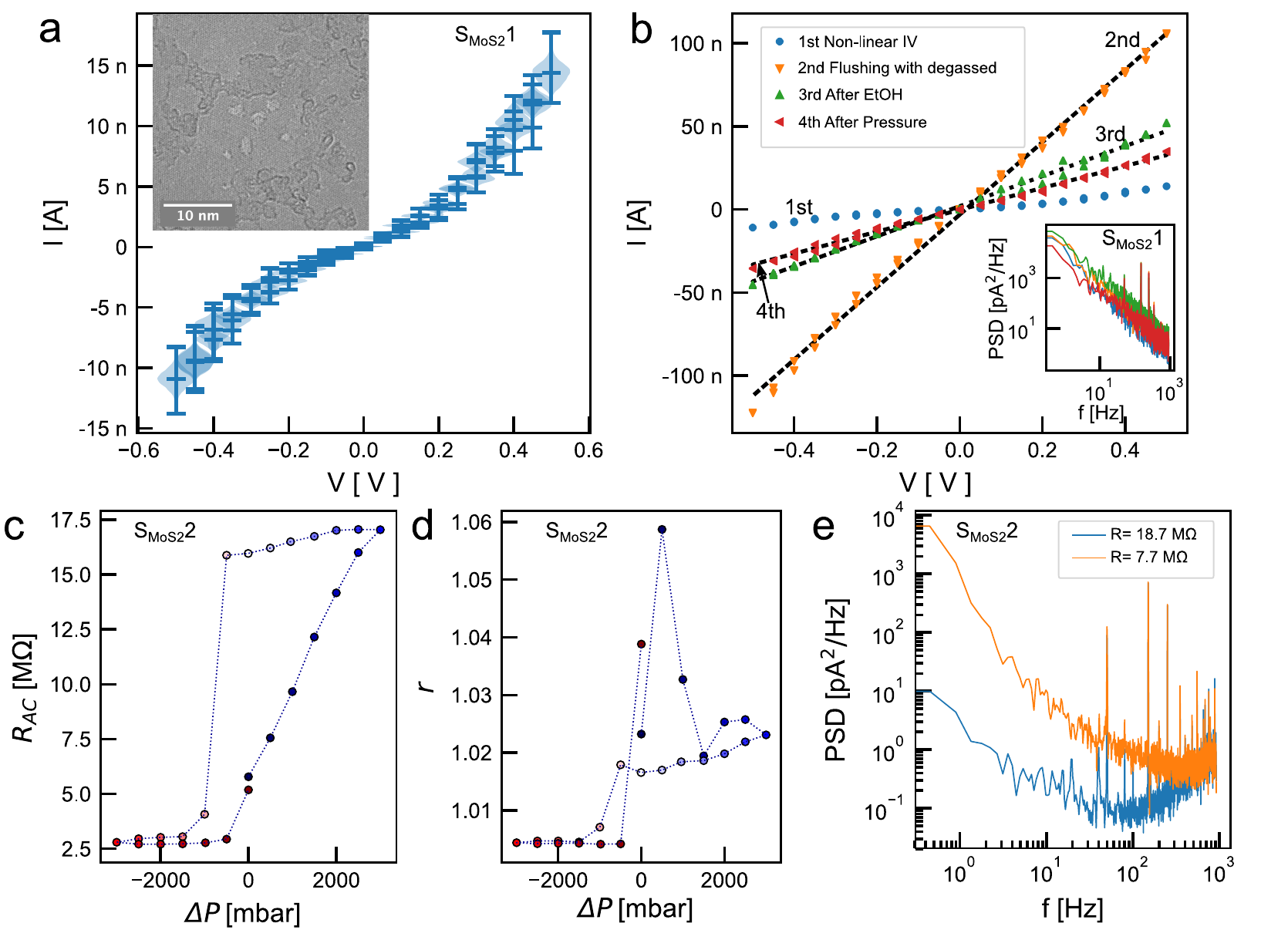}
\caption{\small \textbf{Wetting of nanopores in single layer MoS$_2$ membranes.} a) Stabilized  IV curve of an MoS$_2$ sample (S$_{MoS2}$3) after 1 day of measurements (See SI). The plot bars show the extrema of the current distributions, while the width of the plots represent the proability density for the measured current level at that voltage. Inset is an TEM micrograph of the drilled pores located above the supporting $\approx 75$ nm diameter nanopore in silicon nitride. b) Change of IV curve after wetting with freshly degassed solution from the value shown in panel b (blue circles) to a linearized IV curve (orange, flushing with degassed). The dashed line represents a linear fit giving a resistance of $R=4.8$ M$\Omega$. After a solvent exchange (from 1 M KCl Tris buffered to pH8 to 50\% ethanol 50\% water and back to degassed 1 M KCl Tris buffered to pH8) is performed the resistance of the nanopore increases (dashed line through green points is a linear fit giving a resistance $R=11$ M$\Omega$). After further pressure and voltage is applied the resistance further increases to $R=15.2$ M$\Omega$ ("After pressure", red points). Inset shows the current noise power spectral density for the IV curves in panel b) at a $+0.2$ V bias. c) AC measured resistance $R_{ac}$ versus pressure gradient $\Delta P$ for MoS$_2$ sample S$_{MoS_2}$5. d) AC measured ionic current rectification $r$ versus pressure obtained simultaneously as the measurement in panel c). e) Ionic current power spectral density before and after pressure sweeps for S$_{MoS_2}$5. The noise spectrum is given for a state of resistance $R=18.7$ M$\Omega$, measured at a state before a pressure gradient was used to switch it to the stable ($\Delta P =0$) state in panel c), and $R=7.7$ M$\Omega$ after the pressure sweeps. The noise spectra were obtained under a DC voltage bias of +0.2 V.}
\label{fig:MoS2}
\end{figure*}

Single layer MoS$_2$ supported on the same $\approx$75 nm diameter pores in 20 nm thick silicon nitride as in the previous measurements is used. An example of measurements obtained on a TEM drilled MoS$_2$ nanopore (Sample M1) with several $\approx$2 nm pores are shown on Fig.\ \ref{fig:MoS2}. Only degassed solutions of 1M KCl Tris buffered to pH 8 were used and compression pressures of up to 7 bar were applied while filling the nanopore until the measured capacitance of the membrane $C$ stabilized. The IV response of the nanopore is shown on Fig.\ \ref{fig:MoS2}a. This IV curve shape was persistent for one day of measurements, with fluctuations in the IV curves initially reducing after several hours and the flicker noise increasing. The evolution of the fluctuations is shown in the Supplemental information Sec.\ S6. The shape of the IV curve closely matches the wetting curve in Fig.\ \ref{fig:electrowetting}a with a non-linear activated behaviour. As the TEM imaged size of the pore is above the size where any hydration layer or single ion effects could take place, we attribute this state to a wetting issue. After flushing with degassed solution the pore exhibited a linear IV curve corresponding to a resistance of $R=4.6$ M$\Omega$, or a pore in MoS$_2$ of diameter $d\approx20$ nm (Fig.\ \ref{fig:MoS2}b). Details about MoS$_2$ pore size calculations are provided in the Supplemental information Sec.\ S2. The pore was then probed with a pressure sweep of $\pm 4$ bar to test the stability of the pore resistance. The pore resistance was consistent. At this point the influence of using an alcohol wetting technique was investigated. To do so a solvent exchange procedure\cite{Lou2000} was performed in which a liquid of higher gas capacity (i.e.\ alcohol) is exchanged for one of lower gas capacity (i.e.\ water). A simplified explanation is that in this case the first solvent due to its higher affinity for the hydrophobic surface and higher gas capacity acts as a catalyst for nucleation of gas bubbles. The solution in the chamber was exchanged to 50\% ethanol 50\% water mix, twenty minutes of equilibration time was allowed followed by an exchange back to degassed 1M KCl Tris buffered to pH8. The solvent exchange procedure increased the resistance of the nanopore to $R=11$ M$\Omega$ (calculated pore diameter $d\approx 9$ nm) while retaining a linear IV curve. The flicker noise amplitude $A$ remained the same during all these procedures (Fig.\ \ref{fig:MoS2}b inset), but taking into account the empirical von Hooge relation\cite{Fragasso2019}  $A\sim I^2$ implies that the 2nd curve has the lowest noise as it supports the largest current at a fixed potential $V$ (lowest resistance), and thus the best wetting. The linearity of the curve is in contrast with previous observations where bubbles were shown to produce rectification of the ionic current\cite{Li2015a}. A further application of compression pressure of $7$ bar did not change the resistance, but a pressure gradient sweep in the range $P=\pm 3$ bar changed the resistance to  $R=15.2$ M$\Omega$ or a calculated pore of diameter $d\approx 7$ nm (Fig.\ \ref{fig:electrowetting}b, "After Pressure"). In this case the rectification did not change, as seen in the linearity of the IV curves. Further flushing with degassed solution reduced the resistance to $\approx 2$ M$\Omega$ (close to the value of $1.5$ M$\Omega$ for the supporting nanopore of 75 nm) and was accompanied by a large reduction in the noise level at low frequencies. The potential differences applied on the sample never exceeded $\pm 0.5$ V as higher voltages can cause an electrochemical induced opening of MoS$_2$ pores\cite{Feng2015a}. We hypothesize that the MoS$_2$ was punctured from the start or damaged during one of the flushing procedures and that all measurements except the 2nd (after flushing with degassed solution) were of improper wetting of the pore. Notably, the resistance of this pore was changed, with degassed solutions decreasing it and alcohol wetting increasing it, all at a constant level of flicker noise.

Figure \ref{fig:MoS2}c shows another MoS$_2$ sample (S$_{MoS_S}$2) which was switched between a high ($R_{ac}\approx 16$ M$\Omega$, $d\approx 7$ nm) and low ($R_{ac}\approx 3$ M$\Omega$, $d\approx$ 30 nm) resistance state but with a stable state at no applied pressure ($R_{ac} \approx 6$ M$\Omega$ at $\Delta P=0$). In the high resistance and intermediate state the ionic current rectification was increased, indicating as previously the presence of a bubble. A comparison of the noise before any pressure was applied and after pressure sweeps, indicates that the noise level in the system was increased in the low resistance state. In addition, we performed a streaming potential measurement showing stable fluid flow between two sides of the membrane through the pore before and after this pressure dependent switching (See Supporting information Fig.\ S7).  We interpret this as the bubble changing in size or position, and influencing the noise level at low frequencies. This is consistent with the theory of charge binding/debinding at the bubbles surface being responsible for the low frequency flicker noise\cite{Gravelle2019}, where changing the amount of surface area exposed to the solution and its properties, e.g. liquid-gas or liquid-solid interface, would change the noise spectrum. Additional data on three more samples is provided in the Supplemental information Sec.\ S7, out of which two have been imaged prior to measurements. Fluctuations between linear and nonlinear IV curves are found to be common, and a lower level of the noise power spectral density is not a good measure of bubbles being absent. As with the results from the previous section, the flicker noise can be larger in smaller pores. The gradient pressure clearly shows the ability to change the state of the pore, while compression pressure only changes the state of the pore when a freshly degassed solution is used. One could argue that pressure gradients are damaging the MoS$_2$ and opening up pores. The total force applied on the $75$ nm diameter exposed MoS$_2$ is $\approx1.5$ nN at $3$ bar which is orders of magnitude smaller both than typical forces applied in atomic force microscopy indentation experiments\cite{Liu2014c}, and forces required to delaminate MoS$_2$\cite{Lloyd2017}. We also note that the 20 nm thick supporting silicon nitride membrane can break due to applied pressure gradients at $\approx$5-7 bar only if the membrane area is larger than $\approx$30x30 $\mu$m$^2$.  That being said even if the MoS$_2$ is being enlarged, it does not modify the observations that bubbles are present and can produce both linear and nonlinear IV curves of varying noise levels. The resistance of the pore comes from two series connected terms\cite{Kowalczyk2011}, one being the resistance of the pore interior and the other from the access area. This access area contribution to the resistance and noise is larger for smaller pores\cite{Kowalczyk2011,Fragasso2019}, and we speculate that a bubble obstructing the access area could have a lesser contribution to the noise level than one which has entered the pore interior.

We note that all MoS$_2$ samples presented here show the same pattern. All samples showed apparent resistances at low voltages ($\pm 0.25$ V) indicating either smaller pores than imaged (samples S$_{MoS_2}$1, S$_{MoS_2}$4) or pores in materials which were known to have no pores (S$_{MoS_2}$3). One could naively use this resistance value to infer nanopores in the $\approx$1 nm range if only ionic current measurements were used. At higher voltages they present nonlinear conductivity with the same pattern as seen in the wetting of hydrophobic pores (Fig.\ \ref{fig:bubble}). This nonlinear conductivity disappears after flushing with degassed solution and compressing with pressure. The resistance of the pores can be reduced by applying negative pressure inducing fluid flow from the back side to the front side of the membrane, and increased by applying positive pressure. This is consistent with the bubble being present on the side of the membrane onto which MoS$_2$ was transferred, as negative pressure (from the back side) would then move the bubble away from the pore opening. A recent study on nanopores in supported graphene using atomic force microscopy demonstrated that bubbles in the $\approx$100 nm size range are common and can increase the resistance of the pore and provide a nonlinear signal\cite{Cantley2019}, which we interpret as wetting issues similar to the non linear curves obtained in Fig.\ \ref{fig:bubble}, \ref{fig:electrowetting} and \ref{fig:MoS2}b. The explanation given for this process is that it stems from the use of solvent exchanges from alcohol to aqueous salt solutions, identical to the procedures performed in this study to increase the resistance of the MoS$_2$ pore (Fig.\ \ref{fig:MoS2}b). We conclude by induction that the most probable explanation of such effects is the presence of a bubble in the nm-size range at the nanopore entrance with the supported MoS$_2$ being damaged either before or during wetting.

The possible causes for nanobubbles being stable on the 2D material surface remain to be explained. It is clear that a basic requirement is that the surface of the material be at least mildly hydrophobic\cite{Lohse2015}. 2D materials like graphene, MoS$_2$, h-BN have contact angles which can be larger than the minimum reported for generation of stable nanobubbles on surfaces. This will highly vary on the sample quality and type of supporting surface\cite{Gaur2014,Chow2015,Annamalai2016, Lohse2015}. Nanobubbles have been directly visualized so far only on graphene\cite{Shin2015a,Cantley2019}. Another possibility is contact line pinning of the nanobubble to the surface via any surface defects or contaminants which have been shown to stabilize the bubbles and aid growth\cite{Lohse2015,Fang2016}. Contaminants on 2D material surfaces are a common occurrence in typical methods of transfer from growth to supporting surfaces which involve the use of some form of polymer based stamp\cite{Schneider2010a,Castellanos-Gomez2014,Zomer2011,Graf2019}. The degree of contaminants has been reported to be significantly reduced if a polymer-free transfer method is used\cite{Zhang2016b}. Usually polymers used for transfer involve PMMA or hydrophobic PDMS\cite{Bhattacharya2005}. In the case of PMMA, while a homogeneously smooth coating is not expected to support nanobubble pining, patches are expected to\cite{Agrawal2005}. The MoS$_2$ samples in this study used a PMMA stamp based transfer method which is known to leave hydrocarbon residues\cite{Graf2019}. This is also confirmed by electron micrographs of the samples (TEM images of three MoS$_2$ samples provided in the Supplemental information). Nanobubble nucleation at surfaces has been shown to be possible even at low levels of overgassing (100-120\%) with a temperature change of a few degrees around room temperature inhibiting or promoting nanobubble formation\cite{Seddon2011}. The presence of nano-pits or crevices has been found to increase nanobubble stability\cite{Guo2016a,Wang2017c}, large holes in supported 2D membranes would play this role in the nanopore system. The mechanism of nanobubble nucleation and stability on 2D materials is something which would require further study.

\section{Conclusions}

We show a variety of ionic transport phenomena induced by bubbles or contaminants, and provided a way to control them by using hydraulic pressure gradients between the two sides of a nanopore. This has allowed us to shed light on an important concept for the 2D nanopore community: how hydrophobicity and nanoscale defects or contaminants can enable improper wetting or nanobubbles to imitate other effects. Samples which would normally either be misinterpreted or rejected are found to be plagued by wetting issues. One of the reasons for these wetting issues with 2D materials is the use of alcohol prewetting, a technique prevalent in the solid state nanopore community\cite{Jain2015,Feng2016,Heerema2018,Graf2019}.
This part of a nanopore filling protocol, if combined with hydrophobic or contaminated surfaces, is equivalent to the standard technique of solvent exchange used to nucleate nanobubbles on hydrophobic surfaces\cite{Lohse2015}. We believe that this protocol was carelessly transferred from its use with hydrophilic pore materials to hydrophobic and contaminated 2D materials.  The standard methods of wetting hydrophobic pores which involve high voltages\cite{Beamish2012,Smirnov2011} would damage the material by producing and enlarging surface defects via electrochemical reactions\cite{Feng2015a,Kuan2015}. In this context, we demonstrate that the combination of degassing and applying pressure in a closed air-tight chamber is a suitable method for wetting. Even in the case where complete wetting is not possible due to stability of nanobubbles in degassed solutions\cite{Qian2019}, pressure induced fluid flow has been demonstrated as a useful tool to detect and possibly remove nanobubbles and other contaminants. This is especially relevant in cases where wetting issues reduce an already low yield with complex fabrication protocols, for example the addition of transverse electrodes for DNA translocations\cite{Heerema2018}. DNA translocations have been used as an argument for proper wetting if the current drops match the expected size of the nanopore. If the current drop of the translocating molecule corresponds to the size of the pore it can be a  good indicator of proper pore wetting but if it deviates one should consider among other factors also improper wetting of the pores. A suitable alternative to DNA translocations could be found in streaming measurements\cite{Firnkes2010,Waduge2017}. Streaming is a natural extension to the method presented here and can be performed \textit{in situ} with no modifications to the experimental setup, as demonstrated for the case of MoS$_2$.

Wetting is especially problematic if combined with the study of $\approx$1 nm sized pores in 2D materials, which are expected to have resistances comparable to electrowetting curves of hydrophobic pores at small applied voltages\cite{Smirnov2011}. In the future additional proof needs to be provided of proper wetting of nanopores in 2D materials to corroborate claims of any finite size ion or nonlinear effects\cite{Jain2015,Feng2016} as our work shows that the shape of IV curves can be considerably modified by bubbles and/or unwetted pores which are stable for more than several hours. We have demonstrated that using pressure induced fluid flow can modulate the apparent resistance of a nanopore by moving or changing the size and shape of a pinned gas bubble, but the same approach is also valid for any type of solid obstruction at the pore entrance. Both linear and nonlinear IV curves in the range of $\pm1$ V can be obtained in this way. We also find that the noise level can be reduced by applying pressure as expected by improved wetting\cite{Beamish2012}, but also that it is possible to obtain relative increases in low-frequency noise when the resistance of the pore is increased.  The exact geometry, shape, and size of nanobubbles can influence the level of noise at low frequencies\cite{Fragasso2019,Gravelle2019}, so we argue it will be hard to judge if slight variations in the noise are indicative of the removal of nanobubbles as was previously expected\cite{Smeets2006a,Uram2008,Beamish2012}.  We do notice the presence of current fluctuations between two or more states as reported in works on hydrophobic pores and connected to a wetting/dewetting transition\cite{Smirnov2011,Powell2011} which is also indicative of the presence of bubbles. 

The existence of even the slightest level of ionic current rectification can be an indicator of bad wetting which can modify the resistance of the nanopore. Standard DC methods in use will not always be sufficient. The best approach is to combine this with an \textit{in situ} probe, like laser light irradiation\cite{Smeets2006a} or as demonstrated in our case hydrostatic pressure. Also, as we have seen wetting issues can relax over the time scale of hours, indicating that experiments should be done for longer times using a sealed sample chamber to prove nanopore stability. And in the case of any potentially contaminated and hydrophobic materials (e.g. 2D materials), the technique of solvent exchange should be used with care and preferentially only degassed solutions used.

\section{Methods}

\subsection{Microfluidics chamber}

The microfludic flow-cell was designed to accommodate 5x5mm Si/SiN membrane devices under high working pressure and closed salt solution circulation providing precise pressure control, electrical and thermal insulation. All flow-cell components in contact with the fluid are made from polyether ether ketone (PEEK). Fluid connections to the flow-cell are made with PTFE tubing and connections made with HPLC grade ferrules and fittings. The fluidic pathways are sealed with mechanical shut-off valves. All liquid exchanges are done by flushing liquid through these fluidic connections using Luer-lock syringes. Liquids are degassed by pushing the fluid first through a 4 ml internal volume degassing hose (Biotech Fluidics BT-9000-1549) connected to a vacuum pump at 10 mBar absolute vacuum.  Nitrile O-rings were used to ensure sealing of the chip between the two halves of the chamber. The internal volume of each side of the chamber is $0.125$ ml so we always performed exchanges of solutions by flushing at least $2-3$ mL of liquid trough each side of the chamber, with added pauses to ensure proper mixing even in dead areas without flow.  The flow-cell was cleaned by 20 min sonnication in 70$^\circ$C MiliQ water (18.2 M$\Omega$/cm, 200 nm filtered) to dissolve any remnant salt crystals, then for 20 minutes by sonnicating in isopropanol to remove any greasy residues, then for at least two 20 min sonnications in MiliQ water before drying overnight at 70$^\circ$C. The chamber and the O-rings were treated with oxygen plasma for 30s to make the surface hydrophilic prior to use. Chlorinated Ag/Cl electrodes were sealed using the same fittings and ferrules as the connecting tubing. To ensure there are no air pockets near the electrodes they were partially unscrewed and liquid in the chamber was used to push any air out. 

\subsection{Measurements}

All electrical measurements were done using a Zurich Instruments MFLI lock-in amplifier with the MF-DIG option. Both DC and AC bias was applied using the signal output of the instrument, while the current through the sample was measured using the built in current to voltage converter. All DC (AC demodulator) signals were sampled at 1.83 kHz  (1.6 kHz) and acquired using the MFLI lock-in amplifer. The input noise used by the amplifier depended on the current input range used and was generally bellow 200 fA$ /\sqrt{\text{Hz}}$. All measurements were done inside a Faraday cage. Note that the presence of mains line noise in some of the current spectral power densities was because a short coaxial cable was used between the Faraday cage and MFLI instrument, which coupled with the connection for signal output from the lock-in produced a ground loop. All DC IV curves were recorded in a sweep from $0$ to $+V$, down to $-V$ and back to $0$ to ensure any hysteresis is visible. Thus, for all set voltage values, except the highest and lowest, two points are shown on all figures. Pressure was applied and controlled with $99.99$\% nitrogen using a 7 bar FlowEZ microfluidics pressure controller (Flugient). All interfacing with the measurement instruments was done using a custom made program in LabVIEW. All measurement data was analysed using a custom made script in Python using SciPy signal analysis tools\cite{Virtanen2019}.

All electrical measurements performed while sweeping the pressure were done after the pressure level has stabilized to at least 5\% of the target value. In the case of DC current measurements an additional wait time of 1 s was performed after the pressure settling. In the case of lock-in measurements the wait time was 15 times the lock-in base time plus 2.5 s. The base time constant used was usually 300ms to 1s for an AC signal at 1 Hz, which was a compromise between the speed of measurements and measurement precision.  Pressure measurements and electrical measurements were synchronized only within $\approx 0.25$ s which was taken into account during the analysis. Details on the conversion of raw data to resistance $R_{ac}$ and capacitance $C$, the calculation of the AC rectification factor $r$, as well as the measurement protocol are provided in the SI Sec.\  S2. 

We used 1M KCl with 10 mM Tris buffered to pH 8 for all conductance measurements. All buffers were prepared using MiliQ grade water(18.2 M$\Omega$/cm). The conductivity of all solutions was checked before use with a Mettler-Toledo FiveEasy Plus. For solvent exchange we used a 50\%/50\% mixture of Ethanol and MiliQ water with a measured conductivity of $\approx$10 $\mu$S/cm. All solutions were filtered through a 20 nm filter before use (Whatman Anotop 25 plus). 

\subsection{Supporting membranes and MoS$_2$}

All the measurements provided in the main text were done using in-house fabricated 20 nm thick silicon nitride membranes based on wafers bought from the same supplier. Two additional controls are provided in the SI using commercially bought membranes from NORCADA. Details of the fabrication procedure and samples are provided in Supplemental information Sec.\ S1.

MoS$_2$ was synthesized using a modified growth protocol based on growth promoter spincoating\cite{Cun2019}.  An annealed c-sapphire 2inch wafer was cleaned in IPA/DI and spincoated with sodium molybdate/sodium chloride water mixture (at concentration of 0.03M/0.1M), inserted in the middle of homemade MOCVD hotwall tube furnace and ramped up to 850C under the flow of 210sccm of pure Ar (99.999\%) and ambient pressure. During the growth step, metalorganic precursor (MoCO6, Sigma Aldrich 99.9\%) and diethyl sulphide (C2H6S2, Sigma Aldrich 98.0\%) were supplied from separate bubblers (both at 17C) by an Ar flow of 12 sccm and 3 sccm respectively as well as 4 sccm of H2 and 1 sccm of O2 to improve precursor decomposition, prohibit C contamination and increase the growth yield and quality. After 30min reaction gases were closed and reactor was cooled down naturally. 2D material was transferred to a device using a PMMA based transfer methods reported elsewhere\cite{Graf2019}. MoS$_2$ was imaged and drilled using a FEI TEM Talos with an 80kV electron beam in HRTEM mode. Details are provided in the Supplemental information Sec.\ S1.

\subsection{Supplemental information}

Description of samples used along with supplemental measurements from a total of four hydrophilic nanopores (S1, S2, S3, S4) and five MoS$_2$ devices (S$_{MoS_2}$1, S$_{MoS_2}$2, S$_{MoS_2}$3, S$_{MoS_2}$4, S$_{MoS_2}$5). Description of DC and AC measurements and a discussion of the influence of wetting on the apparent capacitance $C$. A FEM model using coupled Poisson-Nernst-Planck-Stokes equations of the possible scenarios of pore obstruction.  An example of a full time trace of an AC pressure sweep.


\section*{Author contributions}

S.M. designed and built the experimental set-up and designed the study. M.M designed the microfluidic chamber. S.M. and M.M. performed the experiments. A.C. fabricated the devices, transferred and prepared MoS$_2$ devices. M.M. grew the MoS2. A.R. initiated and supervised the research. S.M. analyzed the data, made the FEM model and wrote the manuscript.  All authors provided important suggestions for the experiments, discussed the results, and contributed to the manuscript.

\section*{acknowledgement}
The authors thank J. Gundlach, M. Wanunu, M. Graf, M. Lihter and M. Thakur  for useful discussions. This work was financially supported by the Swiss National Science Foundation (SNSF) Consolidator grant (BIONIC BSCGI0\_157802) and from the European Union's Horizon 2020 research and innovation programme under the Marie Skłodowska-Curie grant agreement No 754462.

%
%
%

\bibliographystyle{apsrev4-1}
\bibliography{/home/smarion/Documents/library.bib}



\end{document}